\documentclass[twocolumn,showpacs,preprintnumbers,amsmath,amssymb,prb]{revtex4}

\usepackage{graphicx}
\usepackage{dcolumn}
\usepackage{bm}

\begin{document}

\title{Time-dependent density-matrix functional theory for biexcitonic phenomena}
\author{V.~Turkowski$^{a}$, C.A.~Ullrich$^{b}$, T.S.~Rahman$^{a}$, and Michael~N.~Leuenberger$^{a}$ }
\altaffiliation{Corresponding author, e-mail address: mleuenbe@mail.ucf.edu}

\affiliation{$^{a}$ Department of Physics and NanoScience and Technology Center,
University of Central Florida, Orlando, FL 32816\\
$^{b}$ Department of Physics and Astronomy, University of Missouri, Columbia, MO 65211}

\date{\today}

\begin{abstract}
We formulate a time-dependent density-matrix functional theory (TDDMFT) approach
for higher-order correlation effects like biexcitons in optical processes in solids
based on the reduced two-particle density-matrix formalism within the normal orbital
representation. A TDDMFT version of the Schr\"odinger equation for biexcitons in terms
of one- and two-body reduced density matrices is derived, which leads to
finite biexcitonic binding energies already
with an adiabatic approximation. Biexcitonic binding energies for several
bulk semiconductors are calculated using a contact biexciton model.
\end{abstract}


\maketitle

\section{Introduction}

The theoretical description of ultrafast processes in modern electronic devices is an
important problem of contemporary condensed matter physics.\cite{1,2}
The entangled role of the fluctuation and correlation effects, especially
in low dimensions, makes such an examination a challenge.
In particular, it is not easy to reproduce correct excitonic and biexcitonic
features in the optical absorption spectra of materials. Besides fundamental
interest such as in 4-wave mixing,\cite{3,4} excitonic and biexcitonic effects
have a variety of practical applications, such as optoelectronic devices,\cite{5}
entangled photon sources,\cite{6} and quantum computing.\cite{7}

The standard approaches, based on the semiconductor Bloch equations (SBEs) \cite{8}
and nonequilibrium Green's function techniques,\cite{9,10} cannot
be easily applied to study higher-order correlation effects  in strongly nonequilibrium situation,
because this requires many-particle correlation functions that depend on many time arguments.\cite{11}
Approaching these problems with time-dependent density-functional theory (TDDFT) \cite{12} looks
promising due to its formal simplicity and the fact that
it in principle includes correlation effects exactly; however, this usually requires going beyond the standard
LDA-GGA approximations.\cite{13} Excitonic effects have been studied with TDDFT in several ways,
including the time-dependent optimized effective potential approach \cite{14} and the combination with the
Bethe-Salpeter method.\cite{1} Unfortunately, these approaches also become
very tedious in the strongly nonequilibrium case.

Recently, we proposed an alternative approach for ultrafast excitonic effects
based on the single-particle density matrix and a TDDFT version of the SBEs.\cite{15,16}
We showed that the effective electron-hole attraction is defined by matrix
elements of the exchange-correlation (XC) kernel $f_{\rm xc}$ with respect to the valence and conduction
band Kohn-Sham single-particle wave functions. Experimentally observed lowest exciton binding energies
can be reproduced in a simple way using the local and long-range XC kernels
\begin{equation}\label{fxcloc}
f_{\rm xc}^{\rm local}({\bf r},{\bf r}')=-A_{0}\delta ({\bf r}-{\bf r}')
\end{equation}
and
\begin{equation}
f_{\rm xc}^{\rm LR}({\bf r},{\bf r}')=-\frac{\alpha}{|{\bf r}-{\bf r}'|},
\label{1}
\end{equation}
where $A_0$ and $\alpha$ can be viewed as adjustable parameters.
The ALDA $f_{\rm xc}$ leads to a too weak electron-hole attraction to produce bound excitons.

In the case of biexcitons, which are correlated double electronic excitations, the problem is much
more complicated. One reason is that multiple excitations in TDDFT require
nonadiabatic XC functionals, and so far there are no simple approximations available.
Another reason is that at first sight it is not clear how to represent biexcitonic  wave
functions in Kohn-Sham TDDFT. In this paper, we formulate and test an alternative
TDDMFT approach for biexcitons based on the natural orbital (NO) representation for the stationary electron
eigenfunctions,\cite{17,18,19} where the multiparticle excited states are naturally related
to the higher-order density matrix elements.

\section{TDDMFT formalism}

The standard TDDFT single-particle Hamiltonian is
\begin{equation} \label{TDDFT_Hamiltonian}
{\hat h}({\bf r},t)=-\frac{\nabla^{2}}{2}+V({\bf r},t)
+
V_{\rm H}[n]({\bf r},t) + V_{\rm xc}[n]({\bf r},t) \:.
\end{equation}
Here, $V({\bf r},t) = V_{\rm nucl}({\bf r})+V_{\rm ext}({\bf r},t)$ is the static potential of the nuclei plus the time-dependent
external perturbing potential. We consider a homogeneous external electric field in dipole approximation,
$V_{\rm ext}({\bf r},t)=-{\bf r}E(t)$,
which implies that the characteristic field frequency is much larger than the level spacing. \cite{21}
$V_{\rm H}({\bf r},t)$ is the Hartree potential, and $V_{\rm xc}({\bf r},t)$
is the time-dependent xc potential, which are both functionals of the time-dependent single-particle density $n({\bf r},t)$.
The Hartree potential is not very important for the description of excitions, but the xc potential is crucial, since it
accounts for the effective electron-hole interaction.\cite{16}

In general, $V_{\rm xc}[n]({\bf r},t)$ has a memory, i.e., it depends on densities at previous times $t'\le t$. The resulting
xc kernel $f_{\rm xc}$, defined as
\begin{equation}
f_{\rm xc}({\bf r},{\bf r}',\omega) = \int d(t-t') e^{i\omega(t-t')}
\left. \frac{\delta V_{\rm xc}({\bf r},t)}{\delta n({\bf r}',t')}\right|_{n_0({\bf r})} \:,
\end{equation}
therefore has in general a frequency dependence. An explicitly frequency-dependent $f_{\rm xc}$ is required for describing
double excitations with linear-response TDDFT.\cite{24} However, to date there are only few approximations for $f_{\rm xc}$ available, and
none of them is particularly suited for the biexcitonic properties in solids we have in mind.
For this reason, we choose a slightly different approach.

To describe the properties of doubly-excited $N$-elec\-tron systems, one can consider the one- and two-electron
density matrices, defined as~\cite{17,18,19}
\begin{eqnarray} \lefteqn{ \hspace{-1cm}
\gamma (x_{1},x_{1}',t)=N\int dx_{2}\int dx_{3}...\int dx_{N}}\nonumber \\
&\times&
\Psi (x_{1},x_{2},...,x_{N},t)\Psi^{*}(x_{1}',x_{2},...,x_{N},t), \label{2}
\end{eqnarray}
\begin{eqnarray} \lefteqn{ \hspace{-1cm}
\Gamma (x_{1},x_{2},x_{1}',x_{2}',t) =
N(N-1)\int dx_{3}\int dx_{4}...\int dx_{N}}\nonumber \\
&\times& \Psi (x_{1},x_{2},...,x_{N},t)\Psi^{*}(x_{1}',x_{2}',...,x_{N},t), \label{3}
\end{eqnarray}
where $\Psi$ is the many-body wave function and $x_{i}=({\bf r}_{i}, s_{i})$
denotes the space coordinate and spin index.
$\gamma (x_{1},x_{1}',t)$ describes the single-particle properties of the system,
such as the charge density $n (x,t)=\gamma (x,x,t)$.
Moreover, all ground-state quantities,
including  $\Gamma_0 (x_{1},x_{2},x_{1}',x_{2}')$, can in principle be obtained from
$\gamma_0 (x_{1},x_{1}')$,  since
there is one-to-one correspondence between $\gamma_0 (x_{1},x_{1}')$
and the ground state many-body wave function $\Psi_0$
(the DMFT generalization of the Hohenberg-Kohn theorem \cite{20}).

Let us now restrict the discussion to two-electron systems and derive
equations of motion for the density matrices. We consider the following effective
two-electron Hamiltonian:
\begin{eqnarray}
{\hat H}({\bf r}_{1},{\bf r}_{2},t)={\hat h}^{\rm ad}({\bf r}_{1},t)+{\hat h}^{\rm ad}({\bf r}_{2},t)+w[n_2]({\bf r}_{1},{\bf r}_{2},t),
\label{4}
\end{eqnarray}
where ${\hat h}^{\rm ad}$ is the TDDFT Hamiltonian (\ref{TDDFT_Hamiltonian}) using an adiabatic approximation for $V_{\rm xc}[n]({\bf r},t)$,
which leads to a frequency-independent xc kernel $f_{\rm xc}^{\rm ad}({\bf r},{\bf r}')$. In this way, excitons can still be
described, since the frequency-dependence of $f_{\rm xc}$ is not essential for the electron-hole interaction. However,
biexcitons (which are correlated two-particle excitations) cannot be captured in the adiabatic approximation. To make up for this,
we introduce, in a somewhat ad-hoc manner, an effective two-particle interaction $w[n_2]({\bf r}_{1},{\bf r}_{2},t)$
which we define as a functional of the two-particle
density $n_2({\bf r}_{1},{\bf r}_{2},t)=\Psi^{*} ({\bf r}_{1},{\bf r}_{2},t)\Psi ({\bf r}_{1},{\bf r}_{2},t)$.
In this way, dynamical screening effects can in principle be accounted for, as done in standard many-body perturbation theory.

In the following we express the two-electron wave-function in terms of the NOs $\chi_{k} ({\bf r})$.
In the singlet case one obtains
$\Psi ({\bf r},{\bf r}',t)=\sum_{k,l} C_{kl}(t)\chi_{k} ({\bf r})\chi_{l}({\bf r}')$, where
$C_{kl}(t)$ is a symmetric matrix, and $k,l$ are the appropriate quantum numbers (band index, momentum, spin etc).
We shall use this matrix in general non-diagonal form for physical insight on the nature of the excitations.
Since the density matrices  $\gamma (x_{1},x_{1}',t)$ and  $\Gamma (x_{1},x_{2},x_{1}',x_{2}',t)$
are defined by the two-electron wave function, they can be expressed in terms of the matrix
elements $C_{kl}(t)$:
\begin{eqnarray}
\gamma (x_{1},x_{1}',t) &=& \sum_{k,l}\gamma_{kl}(t)\chi_{k}(x_{1})\chi_{l}(x_{1}'),
\\
\Gamma (x_{1},x_{2},x_{1}',x_{2}',t) &=&\sum_{klmn}\Gamma_{klmn}(t) \chi_{k}(x_{1})\chi_{l}(x_{2})\nonumber\\
&&\times
\chi_{m}^{*}(x_{1}')\chi_{n}^{*}(x_{2}'),
\label{5}
\end{eqnarray}
where
$\gamma_{kl}(t)=2\sum_{m}C_{km}(t)C_{lm}^{*T}(t)$ and $\Gamma_{klmn}(t)=2C_{kl}(t)C_{mn}^{*}(t)$.
We will soon see that in the two-band approximation the excitonic wave function
is proportional to $\gamma_{{\bf k}_{1}{\bf k}_{2}}^{cv}(t)$, and the biexcitonic one to
$\Gamma_{{\bf k}_{1}{\bf k}_{2}{\bf k}_{3}{\bf k}_{4}}^{ccvv}(t)$,
where $c$ and $v$ stand for the conduction and valence bands, and ${\bf k}_{i}$ is the corresponding
electron and hole momentum. From now on, we shall use superscripts to denote band indices.

The equation of motion for $\gamma_{kl}(t)$ and $\Gamma_{klmn}(t)$ can be obtained via
the equation for $C_{kl}(t)$. From the time-dependent two-electron Schr\"odinger equation
one finds:
\begin{eqnarray}
i\frac{\partial C_{kl}(t)}{\partial t}
&=& \sum_{r}(h_{kr}(t)C_{rl}(t)+C_{kr}(t)h_{rl}(t)) \nonumber \\
&+&\sum_{rs}w_{klrs}C_{rs}(t)
\label{6}
\end{eqnarray}
with the initial condition $C_{kl}(t=0)=\delta_{kl}c_{k}$ and
the matrix elements
\begin{eqnarray}
h_{kr}(t)&=& \int d{\bf r}\chi_{k}^{*}({\bf r}){\hat h}^{\rm ad}({\bf r},t)\chi_{r}({\bf r})\\
w_{klmn}(t)&=&\int d{\bf r}_{1}\int d{\bf r}_{2}\chi_{k}^{*}({\bf r}_{1})\chi_{l}^{*}({\bf r}_{2})
w[n_2]({\bf r}_{1},{\bf r}_{2},t) \nonumber\\
&&
\times \chi_{m}({\bf r}_{1})\chi_{n}({\bf r}_{2}) \:.
\end{eqnarray}
Here and in the following, it is implied that each spatial integration is divided by
the unit cell volume. Equation (\ref{6}) is nonlinear, since the matrix elements, which depend
on the electron density, are functions of $C_{kl}(t)$. From the definitions of
$\gamma_{kl}(t)$ and $\Gamma_{klmn}(t)$ one then obtains the following equations for the one-
and two-particle matrix elements:
\begin{eqnarray}
i\frac{\partial \gamma_{kl}}{\partial t}
&=&\sum_{r}(h_{kr}\gamma_{rl}-\gamma_{kr}h_{rl})
\nonumber \\
&+&\sum_{r,s,m}\left(\Gamma_{krsm}^{*}w_{msrl}^{*}-\Gamma_{krsm}w_{msrl}\right),
\label{7}
\end{eqnarray}
\begin{eqnarray}
i\frac{\partial \Gamma_{klmn}}{\partial t}
&=&
\sum_{r}( h_{kr}\Gamma_{rlmn}+h_{rl}\Gamma_{krmn}
\nonumber \\
&& {}-h_{rm}\Gamma_{klrn} - h_{rn}\Gamma_{klmr})
\nonumber\\
&+&
\sum_{r,s}\left( w_{klrs}\Gamma_{rsmn}
-w_{mnrs}^{*}\Gamma_{klrs}\right).
\label{8}
\end{eqnarray}
An important feature of Eqns. (\ref{7}) and (\ref{8}) is the fact that they are closed, i.e. one does not
need to truncate an infinite hierarchy of equations for higher-order density-matrix elements.
However, keep in mind that this property is only valid for  two-level (two-band) systems.
In the single electron ($w=0$) linearized diagonal approximation for two bands,
one obtains the TDDFT-Wannier equation for the exciton eigenenergies and eigenfunctions from Eq.(7):\cite{16}
\begin{equation} \label{9}
E_{n{\bf q}}^{v}\gamma_{n{\bf k},{\bf q}}^{cv} =
\sum_{{\bf k}'}\left[ \left(\varepsilon_{{\bf k}'+{\bf q}}^{c}-\varepsilon_{\bf k'}^{v}\right)\delta_{\bf k \bf k'}
+F_{{\bf kk}'}\right] \gamma_{n{\bf k}',{\bf q}}^{cv} \:,
\end{equation}
where ${\bf k}$ is the electron momentum, ${\bf q}$ is the sum of the electron and hole momenta
(the exciton momentum) and
\begin{equation}
F_{{\bf kk}'} = 2 \int d{\bf r} \!\int
d {\bf r}'\: \chi_{c{\bf k}}^*({\bf r})\chi_{v{\bf k}}({\bf r}')
f_{\rm xc}({\bf r},{\bf r}')
\chi_{v{\bf k}'}^{*}({\bf r}')\chi_{c{\bf k}'}({\bf r}')
\end{equation}
are the matrix elements for the effective electron-hole attraction.

\section{Two-level model for biexcitons}

The possibility to obtain a biexcitonic state
with the TDDMFT formalism can be already shown for a two-level model with the energy
levels $E_{1}$ and $E_{2}>E_{1}$ and the HOMO-LUMO gap $E_{g}=E_{2}-E_{1}$.
From Eqs.~(\ref{7}) and (\ref{8}) in the lowest (second) order approximation
(by keeping only those matrix elements which contain no more than two indices ``2''),
one obtains the following system of equations after carrying out a Fourier transformation into the frequency domain:
\begin{eqnarray}
(\omega -E_{g}-F^{1})\gamma^{21}-G^{1}\Gamma^{2211}=0,
\nonumber \\
(\omega -2E_{g}-G^{2})\Gamma^{2211}-F^{2}\gamma^{21}=0,
\label{10}
\end{eqnarray}
where $F^{1}=w_{2121}-w_{1111}+w_{2112}+h_{21}^{\gamma}+w_{2111}^{\gamma}$,
$F^{2}=2h_{21}+w_{2212}+w_{2221}+w_{2211}^{\gamma}$,
$G^{1}=h_{12}+w_{2122}-w_{1112}+w_{2111}^{\Gamma}$,
$G^{2}=w_{2222}-w_{1111}+w_{2211}^{\Gamma}$, and
\begin{eqnarray}
h_{ab}^{\gamma} &=&  \int d1 d2\chi_{a}^{*}(1)\chi_{b}(1)
\frac{\delta V_{\rm xc}(1)}{\delta n(2)}
\chi_{2}(2)\chi_{1}^{*}(2),
\\
w_{abcd}^{\gamma} &=&  \int d1 d2 d3
\chi_{a}^{*}(1)\chi_{b}^{*}(2) \frac{\delta w(1,2)}{\delta n(3)}
\nonumber \\
&\times&
\chi_{c}(1)\chi_{d}(2)\chi_{2}(3)\chi_{1}^{*}(3),
\\
w_{abcd}^{\Gamma} &=&  \int d1 d2 d3 d4
\chi_{a}^{*}(1)\chi_{b}^{*}(2)\frac{\delta w(1,2)}{\delta n(3,4)}
\nonumber \\
&\times&
\chi_{c}(1)\chi_{d}(2)\chi_{2}(3)\chi_{2}(4)
\chi_{1}^{*}(3)\chi_{1}^{*}(4).
\end{eqnarray}
All matrix elements are evaluated at the initial (non-perturbed)
densities, and we use the shorthand notation 1 for ${\bf r}_1$, 2 for ${\bf r}_2$, etc.
The solutions of the system of equations (\ref{10}) can be
easily found:
\begin{equation}
\omega =\frac{3E_{g}+F^{1}+G^{2}}{2}\pm \frac{1}{2}\sqrt{(E_{g}-F^{1}+G^{2})^{2}+4G^{1}F^{2}}.
\end{equation}

From the general solution, one can discuss several limiting cases. In particular, for no
correlations ($F^{1}=F^{2}=G^{1}=G^{2}=0$) one gets a trivial solution with one and two
free excited electrons: $\omega_{1}=E_{g}$, $\omega_{2}=2E_{g}$.
In the absence of a density-dependent two-electron potential ($w[n_{2}]({\bf r}_{1},{\bf r}_{1})=G^{2}=0$),
one finds an excitonic state with energy $\omega_{1}=E_{g}-F^{1}=E_{g}-E_{\rm exc}<E_{g}$.
However, the second root $\omega_{2}$ cannot be lower than $2E_{g}-2E_{\rm exc}$, i.e. the linear adiabatic
approximation does not produce a biexciton, as expected. On the other hand,
for nonzero $w[n_{2}]({\bf r}_{1},{\bf r}_{1})$, when $G^{2}<0$ and $|F^{1}|<|G^{2}|$, one can
obtain a biexcitonic level with binding energy $G^{2}-2F^{1}$.

\begin{figure}[t]
\includegraphics[width=7.5cm]{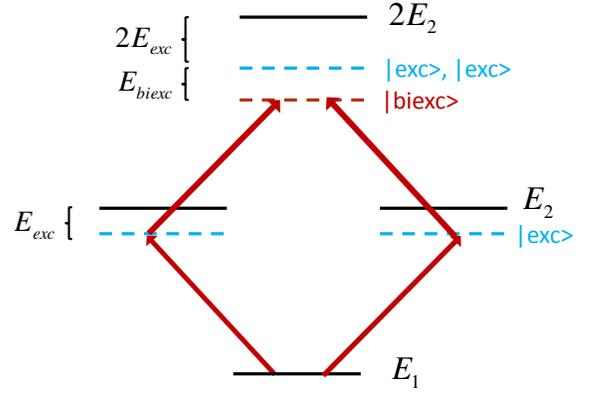}
\caption{\label{fig1} (Color online)
Excitonic bound states in a two-level system with single-particle energies $E_{1}$
and $E_{2}$. $E_{\rm exc}$ and $E_{\rm biexc}$ are the binding energies of excitons and biexcitons, respectively.
A biexcitonic state can be thought of as arising from a two-step process: First, two excitons are created, which then combine to
form a biexciton whose energy is less than the energy of the two individual excitons.}
\end{figure}

Therefore, in order to obtain a biexcitonic state in pure TDDFT [with no $w[n_{2}]({\bf r}_{1},{\bf r}_{1})$]
in the adiabatic approximation, one needs to go to the nonlinear regime and consider the next order terms ($\sim \gamma\Gamma$)
in the equations. To show the possibility of a biexcitonic solution one assumes that the initial state includes
a long-living exciton, $\gamma\Gamma\rightarrow{\bar \gamma}\Gamma$, where ${\bar\gamma}$ is the averaged ``excitonic function''
(see also Ref. \onlinecite{22}, where a possibility to obtain double excitations in the adiabatic TDDFT was considered). Then, the second
(biexcitonic) equation (\ref{10}) acquires an additional term (at $G^{2}=0$), which results in the eigenenergy
$2E_{g}+2(h_{22}^{\gamma}-h_{11}^{\gamma}){\bar \gamma}$ that corresponds to the biexcitonic solution in the case
$(h_{22}^{\gamma}-h_{11}^{\gamma}){\bar \gamma}<0$ and $2E_{g}+2|(h_{22}^{\gamma}-h_{11}^{\gamma}){\bar \gamma}|>2|F^{1}|=2E_{\rm exc}$.
Possible excitations in the two-level case are illustrated in Fig. \ref{fig1}.

\begin{widetext}
\section{The two-band case}

The generalization to the many-electron  case is straightforward. The equation
for the biexciton energies (with zero center-of-mass momentum) in the two-band approximation has the following
form:
\begin{equation} \label{16}
0=
\left[ i\frac{\partial}{\partial t}-\varepsilon_{{\bf k}+{\bf q}}^{c}
-\varepsilon_{{\bf k}'}^{c}+\varepsilon_{\bf k}^{v}
+\varepsilon_{{\bf k}'+{\bf q}}^{v}\right]\Gamma_{{\bf k}+{\bf q},{\bf k}',{\bf k},{\bf k}'+{\bf q}}^{ccvv}
-
\sum_{{\bf k}'}G_{{\bf k}+{\bf q},{\bf k}';{\bar {\bf k}}+{\bar {\bf q}},{\bar {\bf k}}',{\bar {\bf k}},{\bar {\bf k}}'+{\bar {\bf q}}}
\Gamma_{{\bar {\bf k}}+{\bar {\bf q}},{\bar {\bf k}}',{\bar {\bf k}},{\bar {\bf k}}'+{\bar {\bf q}}}^{ccvv} \:,
\end{equation}
where
\begin{eqnarray}
G_{{\bf k}+{\bf q},{\bf k}';{\bar {\bf k}}+{\bar {\bf q}},{\bar {\bf k}}',{\bar {\bf k}},{\bar {\bf k}}'+{\bar {\bf q}}}
&=&
C_{{\bf k}+{\bf q},{\bf k}'}^{cv}
\left( A_{{\bf k}';{\bar {\bf k}}+{\bar {\bf q}},{\bar {\bf k}}',{\bar {\bf k}},{\bar {\bf k}}'+{\bar {\bf q}}}^{1}
+A_{{\bf k}+{\bf q},{\bf k}';{\bar {\bf k}}+{\bar {\bf q}},{\bar {\bf k}}',{\bar {\bf k}},{\bar {\bf k}}'+{\bar {\bf q}}}^{2}
\right)
+A_{{\bf k}+{\bf q},{\bf k}';{\bar {\bf k}}+{\bar {\bf q}},{\bar {\bf k}}',{\bar {\bf k}},{\bar {\bf k}}'+{\bar {\bf q}}}^{3}
\nonumber\\&&
+({\bf k}+{\bf q}\leftrightarrow{\bf k}'),
\\
A_{{\bf k};{\bf k}_{1},{\bf k}_{2},{\bf k}_{3},{\bf k}_{4}}^{1}
&=&
\int d1 d2 d3
\chi_{c{\bf k}}^{*}(1)\chi_{v{\bf k}}^{*}(1)g_{1}(1,2,3)
\chi_{c{\bf k}_1}(2)\chi_{c{\bf k}_2}(3)
\chi_{v{\bf k}_3}(2)^{*}\chi_{v{\bf k}_4}^{*}(3),
\\
A_{{\bf k},{\bf k}';{\bf k}_{1},{\bf k}_{2},{\bf k}_{3},{\bf k}_{4}}^{2} &=&  \int d1 d2 d3 d4
\chi_{c{\bf k}}^{*}(1)\chi_{c{\bf k}'}^{*}(2)\chi_{c{\bf k}}(1)\chi_{v{\bf k}'}(2)
g_{2}(1,2,3,4)\chi_{c{\bf k}_1}(3)\chi_{c{\bf k}_2}(4)
\chi_{v{\bf k}_3}(3)^{*}\chi_{v{\bf k}_4}^{*}(4),
\\
A_{{\bf k},{\bf k}';{\bf k}_{1},{\bf k}_{2},{\bf k}_{3},{\bf k}_{4}}^{3}
&=&\frac{1}{2}A_{{\bf k},{\bf k}';{\bf k}_{1},{\bf k}_{2},{\bf k}_{3},{\bf k}_{4}}^{2}
[\chi_{c{\bf k}}({\bf r}_{1})\rightarrow\chi_{v{\bf k}}({\bf r}_{2})],
\label{18}
\end{eqnarray}
and $g_{1}({\bf r},{\bf r}_{1},{\bf r}_{2})=\frac{\delta V_{xc}({\bf r})}{\delta n({\bf r}_{1},{\bf r}_{1})}$ and
$g_{2}({\bf r}_{1},{\bf r}_{2},{\bf r}_{3},{\bf r}_{4})
=\frac{\delta w({\bf r}_{1},{\bf r}_{2})}{\delta n({\bf r}_{3},{\bf r}_{4})}$
are two-particle density kernels. Similar to the excitonic case, Eq.~(\ref{16}) is the momentum representation
version of the Schr\"odinger equation for two electrons and two holes,\cite{biexcitons} where the matrix elements
$G_{{\bf k}+{\bf q},{\bf k}';{\bar {\bf k}}+{\bar {\bf q}},{\bar {\bf k}}',{\bar {\bf k}},{\bar {\bf k}}'+{\bar {\bf q}}}$
correspond to an integral inter-particle (in general, four-body) interaction.

To solve Eq.~(\ref{16}), we expand the biexcitonic function in terms of the complete set of the excitonic functions
$\gamma_{n,{\bf k},{\bf q}}^{cv}$ with eigenenergies $E_{n,{\bf q}}$ ($n$ is the number of the bound state),
which can be found from the solution of Eq.~(\ref{9}), and antisymmetrize it with respect to interchange of holes
and electrons, in order to satisfy the Pauli principle. Then the biexcitonic functions can be expressed in the following form:
\begin{equation} \label{19}
{\tilde \Gamma}_{{\bf k}+{\bf q},{\bf k}',{\bf k},{\bf k}'+{\bf q}}^{cc'vv'\pm}
=\sum_{n,m}\left[ \gamma_{n,{\bf k}+{\bf q},{\bf q}}^{v}
\gamma_{m,{\bf k}'+{\bf q},-{\bf q}}^{v'}b_{nm,{\bf q}}^{\pm}
\mp
\gamma_{n,{\bf k}',{\bf k}'-{\bf k}}^{v}
\gamma_{m,{\bf k}+{\bf q},{\bf k}-{\bf k}'}^{v'}b_{nm,{\bf k}'-{\bf k}}^{\pm}
\right] ,
\end{equation}
where $\pm$ correspond to two possible states of biexcitons, singlet $(-)$ and triplet $(+)$, with respect
to two-electron spins.\cite{21} Thus the problem is reduced to finding the matrix elements that enter into Eq.~(\ref{16})
and using the orthogonality of the excitonic eigenfunctions, one finds the equation
for the biexcitonic eigenvectors and the corresponding eigenenergies, similar to Eq.~(\ref{9}):
\begin{equation}
\sum_{n',m',{\bf q}'}\left[
\left(\omega -E_{n{\bf q}}-E_{m{\bf q}}\right)\delta_{nn'}\delta_{mm'}\delta_{{\bf q}{\bf q}'}
-H_{nm,n'm',{\bf q}{\bf q}'}^{\pm}
\right] b_{n'm',{\bf q}'}^{\pm}=0,
\label{20}
\end{equation}
where
\begin{eqnarray}
H_{nm,n'm',{\bf q}{\bf q}'}^{\pm}&=&([1\mp {\hat S}]^{-1}{\hat W}^{\pm})_{nm,n'm',{\bf q}{\bf q}'}^{\pm} \:,
\\
{\hat S}_{nm,n'm',{\bf q}{\bf q}'}
&=&\sum_{{\bf k}}\gamma_{n,{\bf k}+{\bf q},{\bf q}}^{v}\gamma_{m,{\bf k}+{\bf q}',-{\bf q}}^{v'}
\gamma_{n',{\bf k}+{\bf q}',{\bf q}'}^{v}\gamma_{m',{\bf k}+{\bf q},-{\bf q}'}^{v'} \:,
\end{eqnarray}
and
\begin{eqnarray}
{\hat W}^{\pm}
&=&
-\delta_{mm'}\delta_{{\bf q}{\bf q}'}
\sum_{{\bf k},{\bf k}'}\gamma_{n,{\bf k}+{\bf q},{\bf q}}^{v}F_{{\bf k},{\bf k}'}
\gamma_{n',{\bf k}+{\bf q}',{\bf q}'}^{v}
-\delta_{nn'}\delta_{{\bf q}{\bf q}'}
\sum_{{\bf k},{\bf k}'}\gamma_{m,{\bf k},-{\bf q}}^{v}F_{{\bf k},{\bf k}'}
\gamma_{m',{\bf k}',-{\bf q}}^{v}
\nonumber \\
&\pm&
\sum_{{\bf k},{\bf k}'}F_{{\bf k},{\bf k}'}^{*}\left(
\gamma_{n,{\bf k}'+{\bf q},{\bf q}}^{v*}\gamma_{m,{\bf k}+{\bf q}',-{\bf q}}^{v'*}
\gamma_{n',{\bf k}+{\bf q}',{\bf q}'}^{v}\gamma_{m',{\bf k}+{\bf q}',-{\bf q}}^{v'}
+\gamma_{m,{\bf k}',-{\bf q}}^{v*}\gamma_{n,{\bf k}-{\bf q}'+{\bf q},{\bf q}}^{v'*}
\gamma_{n',{\bf k},{\bf q}'}^{v}\gamma_{m',{\bf k}-{\bf q}'+{\bf q},-{\bf q}'}^{v'}
\right)
\nonumber \\
&-&\sum_{{\bf k},{\bf k}',{\bar {\bf k}},{\bar {\bf k}}'}
\gamma_{n,{\bf k}+{\bf q},{\bf q}}^{v*}\gamma_{m,{\bf k}',-{\bf q}}^{v'*}
\left[
G_{{\bf k}+{\bf q},{\bf k}';{\bar {\bf k}}+{\bf q}',{\bar {\bf k}}',{\bar {\bf k}},{\bar {\bf k}}'+{\bar {\bf q}}'}
\mp
G_{{\bf k}+{\bf q},{\bf k}';{\bar {\bf k}}',{\bar {\bf k}}+{\bf q}',{\bar {\bf k}},{\bar {\bf k}}'+{\bar {\bf q}}'}
\right]
\gamma_{n',{\bar {\bf k}}+{\bf q}',{\bf q}'}^{v}\gamma_{m',{\bar {\bf k}}',-{\bf q}'}^{v'}.
\label{21}
\end{eqnarray}
\end{widetext}
Equation (\ref{20}) formally resembles equation (\ref{9}) for excitons,  with the exciton eigenenergies
used instead of the bare single-electron energies. The effective exciton-exciton attraction
$H_{nm,n'm',{\bf q}{\bf q}'}^{\pm}$ is defined by the matrix elements of the kernels
$g_{1}({\bf r}_{1},{\bf r}_{2},{\bf r}_{3})$ and $g_{2}({\bf r}_{1},{\bf r}_{2},{\bf r}_{3},{\bf r}_{4})$.

To test the formalism, we consider the simple case of a single biexciton in one-exciton
level approximation, $n=m=1$ (see also Fig. \ref{fig1}). We have obtained the solution of this equation using simple model kernels:
the local kernel (\ref{fxcloc}) to generate the excitonic states, and the following local two-particle kernels for biexcitons:
\begin{equation}\label{g1}
g_1^{\rm local}({\bf r},{\bf r}_{1},{\bf r}_{2})=-C_{0}A_{1}\delta ({\bf r}-{\bf r}_{1})\delta ({\bf r}-{\bf r}_{2})
\end{equation}
[which includes the averaged element $C_{0}$ of the one-electron excited density-matrix component $C^{cv}$,
see Eqns.~(\ref{16})--(\ref{18})] and
\begin{equation}\label{g2}
g_2^{\rm local}({\bf r},{\bf r}',{\bf r}_{1},{\bf r}_{2})
=-A_{2}\delta ({\bf r}-{\bf r}')\delta ({\bf r}-{\bf r}_{1})\delta ({\bf r}-{\bf r}_{2}) \:.
\end{equation}
The kernels (\ref{g1}) and (\ref{g2}) can be viewed as constituting a ``contact biexciton'' model, in analogy with
the contact exciton model defined by the xc kernel (\ref{fxcloc}).\cite{sottile}

Results for the electron eigenenergies and eigenfunctions of several semiconductors were obtained by using the VASP 4.6 code \cite{23}
with GGA-PAW potentials and a 350 eV energy cutoff. We approximated the NO functions by the corresponding
Kohn-Sham single-particle wave functions, which can be considered as a good approximation when the correlations are not too strong.
We find that with the effective local kernels $f_{\rm xc}^{\rm local}$, $g_{1}^{\rm local}$ and $g_{2}^{\rm local}$ one can
reproduce the experimental biexcitonic binding energies with a suitable choice of the parameters $A_0$, $A_1$ and $A_2$ (see Table I).

\begin{table}
\caption{Excitonic and biexcitonic binding energies $E_{b}^{\rm exc}$ and $E_{b}^{\rm biexc}$ for
some semiconductors (in meV), calculated with the TDDMFT formalism.
The parameters $A_0$, $A_1$ and $A_2$ which determine the model interaction kernels (\ref{fxcloc}), (\ref{g1})
and (\ref{g2}) have been adjusted so that the calculations reproduce
the experimental exciton binding energies.} \label{Table}
\begin{ruledtabular}
\begin{tabular}{cccccc}
      & $A_{0} $ & $E_{b}^{\rm exc}$ & $C_{0}A_{1}/\Omega^{2}$ & $A_{2}/\Omega^{3}$ & $E_{b}^{\rm biexc}$\\
 ZnO  &  290             &    60              &       1.82                      &      101                   &  15                \\
 CdS  &  308             &    28              &       0.022                     &        0.64                &  5.7               \\
 CuCl &  20.7            &   190              &       1.97                      &       37.2                 &  32                \\
 CuBr &  20.9            &   110              &       2.0                       &       11.5                 &  25
\end{tabular}
\end{ruledtabular}
\end{table}

Let us briefly discuss how one can in principle find the nonadiabatic kernel $f_{\rm xc}(\omega )$ from the adiabatic
potential $w[n]({\bf r}_{1},{\bf r}_{2})$ which produces biexcitonic states. For this, one can use an approach
similar to the one proposed by Maitra et al. for double excitations in the two-electron case. \cite{24}
Namely, the expression for $f_{\rm xc}(\omega )$ can be obtained by expanding the excited two-electron excited
wave function $\Psi ({\bf r}_{1},{\bf r}_{2},t)$ in terms of quasi-degenerate wave functions of the single-particle
excited and biexcitonic states, and by comparing the eigenenergy equation [for the Hamiltonian (\ref{4}), which
includes $w[n]({\bf r}_{1},{\bf r}_{2})$] with the corresponding TDDFT Casida equation. The single-electron excited
states have to include all states which are close to the biexcitonic one and are well separated from the other states.
A detailed formulation for such a general case will be reported elsewhere.

\section{Conclusion}

In this paper we have formulated a TDDMFT approach to study biexcitonic effects. We have derived the TDDMFT version
of the Schr\"odinger equation for biexcitons in terms of the two-particle density-matrix elements in the two-band
approximation. We have solved this equation in the case of several semiconductors by using phenomenological
two-electron interaction kernels, thereby defining a contact biexciton model. With this model one can reproduce the lowest biexcitonic
binding energies by using proper kernel parameters. To obtain biexcitonic states
within the single-particle TDDFT approach, one would either need to use a frequency-dependent XC kernel or consider the nonlinear regime.
Generalization for the case of bound states with larger number of particles is in principle straightforward.

There are several advantages
of this simplified formalism for biexcitons comparing to other approaches: 1) it can be adapted for use in the real-time domain
in a straightforward manner; 2) physical transparency of the method, in particular the effective TDDFT electron-hole
and exciton-exciton interactions are directly related to the interaction kernels, which may allow one to make
simple estimations of the possibility to produce bound states with given TDDFT kernels; 3) in many cases it may allow
one to construct the non-adiabatic Kohn-Sham XC kernel from the phenomenological adiabatic two-particle density kernel,
which may shed some light on the general requirements on the Kohn-Sham $f_{\rm xc}({\omega})$ necessary to produce
biexcitons and other higher-order coupled states.

Examination of ultrafast processes and higher-order correlation effects, including the excitonic and biexcitonic
transport, in semiconductor nanostructures and organic molecules is underway.

\section{Acknowledgements}

This work was supported in part by DOE-DE-FG02-07ER15842 (V.T. and T.S.R.) and
NSF-ECCS 072551, NSF-ECCS-0901784, AFOSR Grant No. FA9550-09-1-0450 and through the DARPA/MTO
Young Faculty Award HR0011-08-1-0059 (M.N.L.). C.A.U. acknowledges support from NSF Grant No. DMR-0553485.

\end{document}